\documentclass[pre,superscriptaddress,showpacs,twocolumn,floatfix]{revtex4}
\usepackage{color,epsfig}
\usepackage{bm}
\usepackage{amsmath,amsfonts,amssymb,bm}
\usepackage{natbib}

\begin{document}

\title{Quantum Heat Engine in the relativistic limit: The case of a Dirac-particle}
\author{Enrique Mu\~{n}oz}
\affiliation{Facultad de F\'isica, Pontificia Universidad Cat\'olica de Chile, Casilla 306, Santiago 22, Chile.}
\author{Francisco ~J.~Pe\~na}
\affiliation{Instituto de F\'isica, Pontificia Universidad Cat\'olica de Valpara\'iso,
Av. Brasil 2950, Valpara\'iso, Chile.}
\date{\today}

\begin{abstract}
We studied the efficiency of two different schemes for a quantum heat engine, by considering
a single Dirac particle trapped in a one-dimensional potential well as the "working substance". 
The first scheme is a cycle, composed of two adiabatic and two iso-energetic reversible trajectories in 
configuration space. The trajectories are driven by a quasi-static deformation of the potential well due to an external 
applied force. The second scheme is a variant of the former, where iso-energetic trajectories are
replaced by isothermal ones, along which the system is in contact with macroscopic thermostats. This second 
scheme constitutes a quantum analogue of the classical Carnot cycle.
Our expressions, as obtained from the Dirac single-particle spectrum, converge in 
the non-relativistic limit
to some of the existing results in the literature for the Schr\"odinger spectrum. 
\end{abstract}


\maketitle

\section{Introduction}

A classical heat engine consists of a cyclic sequence of reversible transformations
over a "working substance", typically a macroscopic mass of fluid enclosed in a cylinder with a mobile piston at
one end \cite{Fermi_Thermo,Callen}. The two most famous examples are the Otto and Carnot cycles.
In particular, the classical Carnot cycle comprises four stages,  two isothermal and two adiabatic (iso-entropic) ones. Ideal quasi-static and reversible conditions are achieved by assuming that an external 
force, which differs only infinitesimally from the force exerted by the internal pressure of the fluid, is applied to the piston in order to let it move extremely slowly\cite{Fermi_Thermo,Callen}. 
On the other hand, the isothermal trajectories
are performed by bringing the fluid contained by the cylinder into thermal equilibrium
with external reservoirs at temperatures $T_{C} < T_{H}$, respectively.   

A quantum analogue of a heat engine involves a sequence of transformations
(trajectories) in Hilbert's space, where the "working substance" is
of quantum mechanical nature\cite{Bender_02,Bender_Brody_00,Wang_011,Wang_He_012,Quan_06,Arnaud_02,Latifah_011,Quan_Liu_07,Quan_Zhang_05}. One of the simplest conceptual realizations
of this idea is a system composed by a single-particle trapped in a one-dimensional infinite potential well \cite{Wang_He_012,Wang_011,Bender_02,Bender_Brody_00,Latifah_011}.
The different trajectories are driven by a quasi-static deformation
of the potential well, due to the application of an external force. Two different
schemes of this process have been discussed in the literature, in the context
of a non-relativistic particle whose energy eigenstates are determined by
the Schr\"odinger spectrum \cite{Bender_02,Bender_Brody_00,Quan_Liu_07}.
In this paper, we shall revisit these approaches, and we will study the
performance of the corresponding heat engine for a single Dirac particle.
Since Dirac's equation describes the spectra of
relativistic particles, results obtained for this case should reduce
to the corresponding ones from Schr\"odinger's equation in the non-relativistic limit. 
As we shall discuss
below, the transition between the relativistic and non-relativistic regimes
is determined by the ratio $\lambda/L$, with $\lambda = 2\pi\hbar/mc$ 
the Compton wavelength of the particle, and $L$ the width of the potential well. The non-relativistic
limit corresponds to the regime where $\lambda/L \ll 1$, while evidence of the
underlying relativistic nature of the spectrum manifests in terms of finite corrections in powers of $\lambda/L$.
Another limit of theoretical interest is the "ultra-relativistic" case of massless Dirac particles,
where $\lambda/L\rightarrow\infty$. An important realization of this former case in solid state systems 
is provided by conduction electrons in the vicinity of the so called Dirac point in graphene
\cite{Peres010,Castro_Neto09,Munoz010,Munoz012}. 

\section{A Dirac particle trapped in a one-dimensional infinite potential well}

The problem of a Dirac particle in the presence of a one-dimensional, finite
potential well $V(x)$ is expressed by the Dirac Hamiltonian operator \cite{Thaller,Bjorken_Drell}, 
\begin{eqnarray}
\hat{H} = -i\hbar c\bm{\alpha}\cdot\nabla + m c^{2}\hat{\beta} + V(x)\hat{\mathbf{1}}.
\label{eq1}
\end{eqnarray}
Here,
\begin{eqnarray} 
\hat{\alpha}_{i} = \left(\begin{array}{cc}0 & \hat{\sigma}_{i}\\\hat{\sigma}_{i} & 0\end{array} \right), && \hat{\beta} = \left(\begin{array}{cc}I & 0\\ 0 & -I \end{array} \right)\nonumber
\end{eqnarray}
are
Dirac matrices in 4 dimensions, with $\hat{\sigma}_{i}$ the Pauli matrices. The domain
of this operator is ${\mathcal{D}}(\hat{H})=\mathcal{H}$, with ${\mathcal{H}} = L^{2}(\mathbb{R})\oplus
L^{2}(\mathbb{R})\oplus L^{2}(\mathbb{R}) \oplus L^{2}(\mathbb{R}) \equiv L^{2}(\mathbb{R},\mathbb{C}^4)$
the Hilbert space of (complex-valued) 4-component spinors $\hat{\psi}(x)=(\phi_1,\phi_2,\chi_3,\chi_4)$,
where each component $\phi_{i},\chi_{j}\in L^{2}(\mathbb{R})$ is therefore 
a square-integrable function in the unbounded domain $\mathbb{R}$. 
For a {\it{finite}} potential well, of the form $V(x) = V_{0}\Theta(|x|-L/2)$ 
with $0< V_{0} <\infty$,
it is well discussed in the classical literature \cite{Sakurai,Thaller}
that confinement is possible only if the energy $E$
of the particle inside the well is in the interval $|E - V_{0}| < m c^2$, which
corresponds to an exponentially vanishing probability current outside ($x>L/2$ or $x<-L/2$) the confining region. If, on the contrary, the energy is such
that $E - V_{0} <  -mc^2$, then so-called Klein tunneling occurs: 
The particle can tunnel through the barrier with a finite
probability current, but paradoxically with an antiparticle character. This behavior and some of its 
consequences is denominated Klein's paradox\cite{Sakurai,Thaller,Bjorken_Drell}. We remark that this effect occurs when attempting to confine the particle with a {\it{finite}}
potential well.   

The mathematical and physical pictures are rather different when
considering the singular limit of an {\it{infinite}} potential well,
\begin{eqnarray}
V(x) = \left\{ \begin{array}{cc}0\,,& |x|\le L/2\\ +\infty\,, & |x| > L/2  \end{array}\right..
\label{eq2}
\end{eqnarray}

The singular character of the infinite potential well, exactly as in the more familiar
Schr\"odinger case \cite{Carreau_90}, requires a different mathematical statement of the problem: one needs
to define a self-adjoint extension \cite{Thaller,Carreau_90,Alonso_97,Alberto_96} of the {\it{free}} particle Hamiltonian
\begin{eqnarray}
\hat{H}_{0} = -i\hbar c\bm{\alpha}\cdot\nabla + m c^{2}\hat{\beta},
\label{eq3}
\end{eqnarray}
whose domain 
$\mathcal{D}(\hat{H}_{0})\subset \mathcal{H}_{\Omega}$
is a dense proper subset of the Hilbert space 
$\mathcal{H}_{\Omega}=L^{2}(\Omega)\oplus L^{2}(\Omega) \oplus L^{2}(\Omega)\oplus L^{2}(\Omega)\equiv L^{2}(\Omega,\mathbb{C}^4)$ of square-integrable (complex-valued) 4-component spinors in the closed
interval $x\in \Omega = [-L/2,L/2]$. In general, the domain of $\hat{H}_{0}$ and its adjoint
$\hat{H}_{0}^{\dagger}$ verify  $\mathcal{D}(\hat{H}_{0})\subseteq \mathcal{D}(\hat{H}_{0}^{\dagger})$ \cite{Thaller}. However,
physics requires for $\hat{H}_{0}$ to be self-adjoint. The self-adjoint extension
is obtained by imposing appropriate boundary conditions \cite{Thaller,Carreau_90,Alonso_97,Alberto_96} on the spinors 
at the boundary $\partial \Omega$ of the finite domain $\Omega$, as discussed in detail in Appendix A. In
particular, the condition
of simultaneous vanishing of the four components of the spinor at the boundaries $x=\pm L/2$
is not compatible with self-adjointness and, moreover, leads to the trivial
null solution $\hat{\psi}(x)=0$ $\forall x \in \Omega$. Instead, as shown in Appendix A, 
the mathematical condition for self-adjointness
corresponds to a vanishing probability current at the
boundary $\partial\Omega$ of the domain $\Omega=[-L/2,L/2]$, $j^{1}(x=\pm L/2)=0$, with
$j^{1}(x) = c \hat{\psi}^{\dagger}(x)\hat{\alpha}_{1}\hat{\psi}(x)$. The physical interpretation
of this mathematical condition is rather obvious: the particle is indeed "trapped" inside
the infinite potential well, since there is a zero probability current through
the boundary walls. This approach has been used in the past, for instance, to model confinement of hadrons in finite regions
of space \cite{Chodos2_74,Chodos_74}.
Summarizing, the eigenvalue problem for the self-adjoint extension
of the free Dirac Hamiltonian Eq.(\ref{eq3}) representing particles
"trapped" inside the infinite potential well is given by
\begin{eqnarray}
\hat{H}_{0}\hat{\psi} = E\hat{\psi},
\label{eq3b}
\end{eqnarray}
subject to the boundary conditions
\begin{eqnarray}
j^{1}(x=\pm L/2) = 0.
\label{eq3c}
\end{eqnarray}
As shown explicitly in Appendix A,
there is a whole family of eigenfunctions of Eq.(\ref{eq3b}) in $\Omega = [-L/2,L/2]$, 
satisfying the boundary conditions Eq.(\ref{eq3c}). This has been investigated for instance in \cite{Alonso_97,Alberto_96}.

\begin{figure}[tbp]
\centering
\epsfig{file=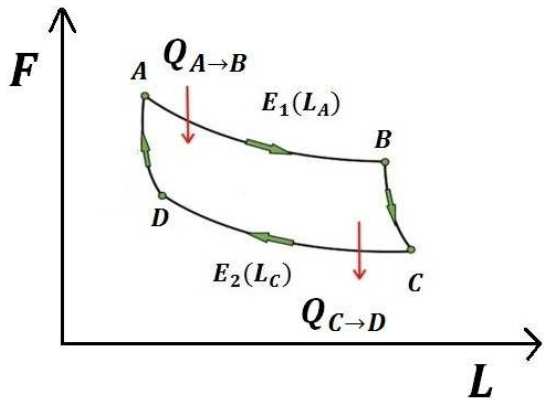,width=0.9\columnwidth,clip=}
\caption{(Color online)Pictorial description of the Force versus size of the potential well, for the Isoenergetic cycle described in the text.
\label{fig1}
}
\end{figure}

On the other hand, a fundamental discrete
symmetry of the Dirac Hamiltonian is its invariance under parity \cite{Sakurai,Bjorken_Drell,Thaller} ($\hat{P}$: $x\rightarrow -x$, $p_{x}\rightarrow -p_{x}$), $[\hat{H}_{0},\hat{P}]=0$, which corresponds to a mirror spatial reflection by leaving the spin direction invariant. 
It is straightforward to show that 
under parity, the spinor transforms as \cite{Bjorken_Drell,Sakurai,Thaller} 
$\hat{P}\hat{\psi}(x)\hat{P}^{-1} = e^{i\phi}\hat{\beta}\hat{\psi}(-x)$. 
On the other hand, the probability density defined as $\rho(x) = \hat{\psi}^{\dagger}(x)\hat{\psi}(x) = c^{-1}j^{0}(x)$ is the time-component of the
Lorentz 4-vector $j^{\mu} = c\bar{\psi}(x)\gamma^{\mu}\hat{\psi}(x)$, with $\bar{\psi}(x) \equiv \gamma^{0}\hat{\psi}^{\dagger}(x)$. Here, the explicit covariant notation $\gamma^{i}=\hat{\beta}\hat{\alpha}_{i}$ (i=1,2,3), $\gamma^{0} = \hat{\beta}$ was invoked. Under parity,
the space components of a Lorentz 4-vector invert sign ($j^{i}\rightarrow -j^{i}$, i = 1,2,3), whereas the time component $j^{0}$ remains invariant \cite{Bjorken_Drell,Sakurai,Thaller}, and hence $\rho(-x) = \rho(x)$. 
For the physically acceptable eigenfunctions of the self-adjoint extension
of the Dirac Hamiltonian Eqs.(\ref{eq3b}),(\ref{eq3c}), we thus demand
that this symmetry is satisfied 
$\forall x \in \Omega = [-L/2,L/2]$. As shown in detail in Appendix A, this leads to 
a quantization of the wavenumbers, $k \equiv k_{n} = n\pi/L$,
for $n \in \mathbb{N}$. The spinor-eigenfunctions
obtained from this analysis are
\begin{eqnarray}
\hat{\psi}_{n}(x) = A\left(\begin{array}{c} \sin(n\pi (x-L/2)/L)\\0\\0\\-\frac{i n\lambda/(2L)}{\sqrt{1 + n^{2}(\lambda/2L)^{2}}}\cos(n\pi (x-L/2)/L)\end{array}\right),
\label{eq4}
\end{eqnarray}
with associated discrete energy eigenvalues,
\begin{eqnarray}
E_{n}^{D}(L) = m c^{2}\left(\sqrt{1 + \left(n\lambda/2L \right)^{2}}-1\right).
\label{eq5}
\end{eqnarray}
Here, we have subtracted the rest energy, and $\lambda = 2\pi\hbar/(mc)$ is 
the Compton wavelength.
The positive sign corresponds to the particle solution \cite{Bjorken_Drell}. 

The spectrum predicted by Eq.(\ref{eq5}) ,
as well as the probability density obtained
from Eq.(\ref{eq4}),
can be compared with the corresponding Schr\"odinger problem,
whose eigenvalues are given by
\begin{eqnarray}
E_{n}^{S}(L) = \frac{n^{2}\pi^{2}\hbar^{2}}{2 m L^{2}}.
\label{eq6}
\end{eqnarray}
Here, a single eigenvalue for the energy is obtained and, moreover, it scales as $n^{2}$, in contrast with the
Dirac particle case where a richer scaling with $n$ is observed.
In particular, in the regime
$\lambda/L \ll 1$, we have:
\begin{eqnarray}
E_{n}^{D}(L) \rightarrow \frac{m c^{2}}{2}\left(n\lambda/2L \right)^{2} =  E_{n}^{S}(L),
\label{eq7}
\end{eqnarray}
corresponding to the non-relativistic limit. Beyond this regime, relativistic corrections
depending on the finite ratio $\lambda/L$ are observed.

Regarding the probability density, in the corresponding Schr\"odinger problem one
obtains $\rho^{S}(x)\propto \sin^{2}(n\pi(x-L/2)/L)$
for $x \in \Omega = [-L/2,L/2]$. It is straightforward to verify that,
in the non-relativistic limit $\lambda/L \ll 1$, the "small" component of the spinor in Eq.(\ref{eq4}) becomes negligible, and thus
the probability density converges to
the Schr\"odinger case, $\rho(x)=\hat{\psi}^{\dagger}(x)\hat{\psi}(x)\rightarrow |A|^{2}\sin^{2}(n\pi(x-L/2)/L)$.

Another interesting limit of Eq.(\ref{eq5}) corresponds to a massless
Dirac particle with $\lambda\rightarrow\infty$, where the spectrum reduces to the expression
\begin{eqnarray}
\left.E_{n}^{D}(L)\right|_{m=0} = \frac{n\pi\hbar c}{L}.
\label{eq8}
\end{eqnarray}
This situation may be of interest in graphene systems, where conduction electrons
in the vicinity of the so called Dirac point can be described as effective massless
chiral particles, satisfying Dirac's equation in two dimensions\cite{Peres010,Castro_Neto09,Munoz010,Munoz012}.

\section{A single-particle Quantum heat Engine}

As the "working substance" for a quantum heat engine, let us consider a statistical ensemble 
of copies of a single-particle system, where each copy may be in any of the possible different energy eigenstates. 
We therefore say that the single-particle system
is in a statistically mixed quantum state\cite{vonNeumann}. The 
corresponding density
matrix operator is $\hat{\rho} = \sum_{n}p_{n}(L)|\psi_{n}(L)\rangle\langle\psi_{n}(L)|$, with $|\psi_{n}(L)\rangle$
an eigenstate of the single-particle Hamiltonian Eq.(\ref{eq3}), corresponding to the spinors
defined by Eq.(\ref{eq4}). This density matrix operator
is stationary, since in the absence of an external perturbation\cite{vonNeumann} $i\hbar\partial_{t}\hat{\rho} = [\hat{H},\hat{\rho}] = 0$. Here, the coefficient $0 \le p_{n}(L)\le 1$
represents the probability for the system, within the statistical ensemble, to be in the particular state
$|\psi_{n}(L)\rangle$. Therefore, the $\{p_{n}(L)\}$ satisfy the normalization condition
\begin{eqnarray}
{\rm{Tr}}\hat{\rho} = \sum_{n}p_{n}(L) = 1.
\label{eq9}
\end{eqnarray}

In the context of Quantum Statistical Mechanics, entropy is defined according
to von Neumann \cite{vonNeumann,Tolman} as $S = -k_{B}{\rm{Tr}}\hat{\rho}\ln\hat{\rho}$. 
Since in the energy eigenbasis the equilibrium density matrix operator is diagonal, the entropy reduces to the explicit expression
\begin{eqnarray}
S(L) = -k_{B}\sum_{n}p_{n}(L)\ln\left(p_{n}(L)\right).
\label{eq10}
\end{eqnarray}
In our notation, we emphasize explicitly the dependence of the energy eigenstates $\{|\psi_{n}(L)\rangle\}$, as well as the probability coefficients $\{p_{n}(L)\}$,
on
the width of the potential well $L$.
\begin{figure}[tbp]
\centering
\epsfig{file=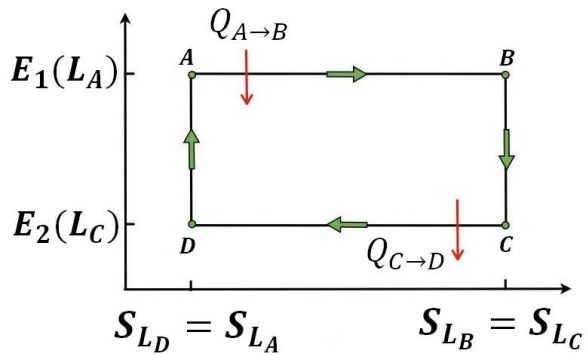,width=0.9\columnwidth,clip=}
\caption{(Color online)Pictorial description of the Energy versus Entropy (von Neumann) for
the Isoenergetic cycle.
\label{fig2}
}
\end{figure}
The ensemble-average energy of the quantum single-particle system is
\begin{eqnarray}
E \equiv \langle \hat{H} \rangle = {\rm{Tr}}(\hat{\rho}\hat{H}) = \sum_{n}p_{n}(L) E_{n}(L).
\label{eq11}
\end{eqnarray} 

For the statistical ensemble just defined, we conceive two different schemes
for a quantum analogue of a thermodynamic heat engine. The first one, that we shall refer to as the Isoenergetic cycle, 
consists on four stages of reversible trajectories: two iso-entropic
and two iso-energetic ones, as originally proposed
by Bender et al.\cite{Bender_02,Bender_Brody_00} in the context of a Schr\"odinger particle. 
During the iso-energetic trajectories, the ensemble-average
energy Eq.(\ref{eq11})
is conserved, while during the iso-entropic 
ones, the von Neumann entropy defined by Eq.(\ref{eq10}) remains constant. We distinguish
this first scheme from the quantum Carnot cycle to be discussed next, where
the iso-energetic trajectories in Hilbert's space are replaced by
isothermal processes. During these stages, the system is brought into thermal equilibrium with
macroscopic reservoirs at temperatures $T_{C} \le T_{H}$, respectively.

\section{The Isoenergetic Cycle}

The Isoenergetic cycle, a scheme for a quantum heat-engine originally proposed by Bender et al. \cite{Bender_02,Bender_Brody_00} in the context of a single Schr\"odinger particle,
is composed by two isoentropic and two isoenergetic trajectories. In particular, during the
isoenergetic trajectories, the "working substance" must exchange energy
with an energy reservoir \cite{Wang_011,Wang_He_012}. A possible practical realization of
this cycle was proposed 
in the context of non-relativistic particles in Ref.\cite{Wang_011}, where the working substance
exchanges energy with single-mode radiation in a cavity, which acts as an energy reservoir. 

The system trajectories in Hilbert's space are assumed to be driven by reversible quasi-static processes, 
in which the walls of the
potential well are deformed "very slowly" by an applied external force, such that the distance $L$ is modified
accordingly. 
Along these trajectories,
the total change in the ensemble average energy of the system is given by
\begin{eqnarray}
d E &=& \sum_{n}p_{n}(L) dE_{n}(L) + \sum_{n} dp_{n}(L) E_{n}(L)\nonumber\\
&=&\left(\delta  E \right)_{\{p_{n}(L)\} = cnt.} 
+ \left(\delta  E  \right)_{\{E_{n}(L)\} = cnt.}
\label{eq12}
\end{eqnarray}
The first term in Eq.(\ref{eq12}) represents the total energy change due to an iso-entropic process, whereas
the second term represents a trajectory where the energy spectrum remains rigid.

Let us consider first the iso-entropic process, where $\{p_{n}(L)\} = cnt$. We remark
that this represents a strong sufficient condition for the entropy to remain
constant along the trajectory, but is not a necessary one\cite{note1}.
Under quasi-static conditions, the external force driving the change in the width of the potential well
is equal to the internal "pressure" of the one-dimensional system, $F = -(\partial E/\partial L)_{S}$.     
Therefore, the work performed by the system against the external force, when
the width of the potential well expands from $L=L_{1}$ to $L=L_{2}$, is given by 
\begin{eqnarray}
W_{1 \rightarrow 2} &=& \int_{L_{1}}^{L_{2}}dL\left(\frac{\partial  E }{\partial L}\right)_{\{p_{n}(L)\} =\{p_{n}(L_{1})\}= cnt.}\nonumber\\
&=& \sum_{n=1}^{\infty}p_{n}(L_{1}) \left[E_{n}(L_{2}) - E_{n}(L_{1})\right].
\label{eq13}
\end{eqnarray}

For the case of a Dirac particle, the work performed under
iso-entropic conditions is given after Eq.(\ref{eq13}) and Eq.(\ref{eq5}) by
\begin{eqnarray}
W_{1 \rightarrow 2} &=& m c^{2} \sum_{n=1}^{\infty}p_{n}(L_{1})\left(
\sqrt{1 + \left(n\lambda/2L_{2} \right)^{2} }\right.\nonumber\\
&&\left.-
\sqrt{1 + \left(n\lambda/2L_{1} \right)^{2}}
\right).
\label{eq14}
\end{eqnarray}
Notice that our sign convention is such that, 
for an expansion process $L_{2} > L_{1}$, the work performed by the
system is negative\cite{Callen}, indicating that the ensemble-averaged energy is decreasing during expansion,
as in a classical ideal gas.

Let us now consider an iso-energetic process, that is, a trajectory in Hilbert space defined
by the equation
$d E = 0$. The solution to this equation, for $L \in [L_{1},L_{2}]$, is given by the path
\begin{eqnarray}
\sum_{n=1}^{\infty} p_{n}(L)E_{n}(L)  = \sum_{n=1}^{\infty} p_{n}(L_{1})E_{n}(L_{1}), 
\label{eq15}
\end{eqnarray}
along with the normalization condition Eq.(\ref{eq9}). 
Clearly, by definition an iso-energetic process satisfies
\begin{eqnarray}
d E  =  \delta W_{1\rightarrow 2} + \delta Q_{1\rightarrow 2} = 0,
\label{eq16}
\end{eqnarray}
with $\delta W_{1\rightarrow 2} \equiv \left(\delta  E  \right)_{\{p_{n}(L)\} = cnt.}$ and
$\delta Q_{1\rightarrow 2} \equiv \left(\delta  E  \right)_{\{E_{n}(L)\} = cnt.}$.
In the former equation, the integral along the trajectory $L_{1}\rightarrow L_{2}$ gives
\begin{eqnarray}
\Delta  E = W_{1\rightarrow 2} + Q_{1\rightarrow 2} = 0.
\label{eq17}
\end{eqnarray}
The first term $W_{1\rightarrow 2}$ corresponds to the mechanical work performed by the system against the external force, which drives the change
in the width of the potential well at constant energy. The second term $Q_{1\rightarrow 2} = -W_{1\rightarrow 2}$ corresponds
to the amount of energy exchanged by the system 
with the environment, in order to rearrange its internal level occupation. This equation
is in analogy with the first law of thermodynamics for macroscopic systems, 
when considering a reversible process over a classical ideal gas which
is being compressed/expanded at constant internal energy conditions. The first term has a precise correspondence with the mechanical work for expansion/compression,
whereas the second is in correspondence with
the heat exchanged by the gas with the environment in order to satisfy total energy conservation. 

\begin{figure}[tbp]
\centering
\epsfig{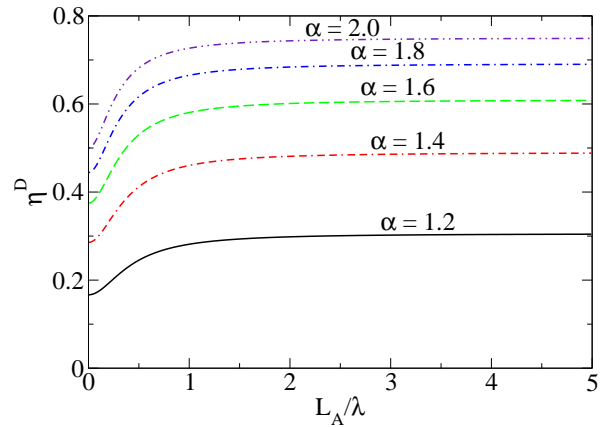}
\caption{(Color online)Efficiency of the Isoenergetic Cycle, calculated after Eq.(\ref{eq28}),
as a function of the expansion parameter $\alpha = L_{C}/L_{B}$.
\label{fig3}
}
\end{figure}

According to the previous analysis, the heat exchanged by the system with the environment along the iso-energetic process is given by
\begin{eqnarray}
Q_{1\rightarrow 2} = \sum_{n=1}^{\infty}\int_{L_{1}}^{L_{2}} E_{n}(L)\frac{dp_{n}(L)}{dL}\,dL.
\label{eq18}
\end{eqnarray}
Evidently, Eq.(\ref{eq15}) combined with the normalization condition Eq.(\ref{eq9}) are not enough to uniquely define the coefficients $p_{n}(L)$
along the iso-energetic trajectory. An exception is the case 
when the energy scale of all the processes involved is such that
only transitions between the ground state ($n=1$) and the first excited state ($n=2$)
are possible. In this effective two-level spectrum, 
combining Eq.(\ref{eq15}) with the normalization condition Eq.(\ref{eq9}), the trajectory for the iso-energetic process is described by the following relation
\begin{eqnarray}
p_{1}(L) &=& \frac{E_{2}(L_{1}) - E_{2}(L)}{E_{1}(L) - E_{2}(L)} + \frac{E_{1}(L_{1}) - E_{2}(L_{1})}{E_{1}(L) - E_{2}(L)}p_{1}(L_{1}),\nonumber\\
\label{eq19}
\end{eqnarray}
with $p_{2}(L) = 1- p_{1}(L)$ after the normalization condition Eq.(\ref{eq9}).
The heat exchanged by the system with the environment during an iso-enegetic trajectory connecting the initial and final states $L_{1}\rightarrow L_{2}$,
for the case of a Dirac particle,
is given by the expression
\begin{eqnarray}
&&-Q_{1\rightarrow 2} = \left[E_{2}^{D}(L_{1}) + \left(E_{1}^{D}(L_{1}) - E_{2}^{D}(L_{1})\right
)\right.\nonumber\\
&&\left.\times p_{1}(L_{1})\right]\ln\left[\frac{E_{1}^{D}(L_{2})-E_{2}^{D}(L_{2})}{E_{1}^{D}(L_{1})-E_{2}^{D}(L_{1})}\right]+2(m c^{2})\ln\left(\frac{L_{2}}{L_{1}}\right)\nonumber\\
&&+ (m c^{2})
\ln\left[\frac{m c^{2}+E_{2}^{D}(L_{2})}{m c^{2} + E_{2}^{D}(L_{1})}\frac{m c^{2}+E_{1}^{D}(L_{2})}{m c^{2} + E_{1}^{D}(L_{1})}\right]
\label{eq20}
\end{eqnarray}
where $E_{n}^{D}(L)$ was defined in Eq.(\ref{eq5}).

For the effective two-level system previously described, we shall conceive a cycle, as depicted in Figure 1, which starts in the ground state with $p_{1}(L_{A}) = 1$. Then, the system experiences an iso-energetic expansion from $L_{A}\rightarrow L_{B}>L_{A}$. Then, it experiences an iso-entropic expansion from $L_{B}\rightarrow L_{C}>L_{B}$, then an iso-energetic compression $L_{C}\rightarrow L_{D}<L_{C}$, and finally it goes back to the initial ground state through an iso-entropic compression $L_{D}\rightarrow L_{A}$. 

We shall assume that
the final state after the iso-energetic process $L_{A}\rightarrow L_{B}$ corresponds to maximal expansion, that is, the system ends completely localized in the excited state $n=2$. In this condition, Eq.(\ref{eq19}) reduces to
\begin{eqnarray}
p_{1}(L_{B}) = 0,\,\,\,\,p_{2}(L_{B}) = 1.
\label{eq21}
\end{eqnarray}
The condition of total energy conservation between the initial and final states connected through an iso-energetic process, for maximal expansion, leads to the equation
\begin{eqnarray}
p_{1}(L_{A})E_{1}(L_{A}) = p_{2}(L_{B})E_{2}(L_{B}),
\label{eq22}
\end{eqnarray}
where $p_{1}(L_{A}) = p_{2}(L_{B}) = 1$ for maximal expansion. Therefore Eq.(\ref{eq22}),
given the Dirac spectrum Eq.(\ref{eq5}),
implies that $L_{B}/L_{A} =  2$.

The heat released to the environment along this first stage of the cycle is calculated after Eq.(\ref{eq20}),
\begin{eqnarray}
-Q_{A\rightarrow B}&=& E_{1}^{D}(L_{A})\ln\left[\frac{E_{2}^{D}(L_{A}) - E_{1}^{D}(L_{A})}{E_{2}^{D}(2L_{A})-E_{1}^{D}(2L_{A})}\right]
\nonumber\\
&+& m c^{2} \ln\left[\frac{1}{4}\left\{\frac{mc^{2} + E_{2}^{D}(L_{A})}{mc^{2}
+ E_{1}^{D}(2 L_{A})}\right\}\right]. 
\label{eq23}
\end{eqnarray}

The next process is an iso-entropic expansion, characterized by the
condition $p_{2}(L_{B}) = p_{2}(L_{C}) = 1$. We shall define
the expansion parameter $\alpha \equiv L_{C}/L_{B} > 1$. The work
performed during this stage, with $L_{B} = 2L_{A}$ (as discussed before), is 
calculated from Eq.(\ref{eq14}),
\begin{eqnarray}
W_{B\rightarrow C} &=& m c^{2}\left[\sqrt{1 + \left(\frac{\lambda}{2 L_{A}} \right)^{2}}
-  \sqrt{1 + \left(\frac{\lambda}{2\alpha L_{A}} \right)^{2}}\right].\nonumber\\
\label{eq24}
\end{eqnarray}

The cycle continues with a maximal compression process from $L_{C} = 2 \alpha L_{A}$
to $L_{D} = \alpha L_{A}$ under iso-energetic conditions. The energy conservation condition is in this case
similar to Eq.(\ref{eq22}), with $p_{2}(L_{C}) = p_{1}(L_{D}) = 1$. The heat exchanged
by the system with the environment along this process, applying Eq.(\ref{eq20}), is given by the expression
\begin{eqnarray}
-Q_{C\rightarrow D}
&=& E_{2}^{D}(2\alpha L_{A})\ln\left[\frac{E_{2}^{D}(\alpha L_{A}) - E_{1}^{D}(\alpha L_{A})}{E_{2}^{D}(2\alpha L_{A})-E_{1}^{D}(2\alpha L_{A})}\right]
\nonumber\\
&&+ m c^{2} \ln\left[\frac{1}{4}\left\{\frac{mc^{2} + E_{2}^{D}(\alpha L_{A})}{mc^{2}+ E_{1}^{D}(2\alpha L_{A})}\right\}\right] 
\label{eq25}
\end{eqnarray}
where $E_{n}^{D}(L)$ was defined in Eq.(\ref{eq5}).
The last path along the cycle is an adiabatic process, which returns the system to its initial
ground state with $p_{1}(L_{D}) = p_{1}(L_{A}) = 1$. The work performed during
this final stage, as obtained by applying Eq.(\ref{eq14}), is given by
\begin{eqnarray}
W_{D\rightarrow A} &=&  m c^{2}\left[\sqrt{1 + \left(\frac{\lambda}{2\alpha L_{A}} \right)^{2}}
- \sqrt{1 + \left(\frac{\lambda}{2 L_{A}} \right)^{2}}\right].\nonumber\\
\label{eq26}
\end{eqnarray}
It is interesting to check that the work along the two iso-entropic trajectories cancels, that is
$W_{B\rightarrow C} + W_{D\rightarrow A} = 0$. Therefore, the efficiency of the cycle is defined by the ratio
\begin{eqnarray}
\eta^{D} = 1 - \frac{Q_{C\rightarrow D}}{Q_{A\rightarrow B}}.
\label{eq27}
\end{eqnarray}
When substituting the corresponding expressions from Eq.(\ref{eq23}) and Eq.(\ref{eq25}) into Eq.(\ref{eq27}),
we obtain the explicit analytical expression
\begin{eqnarray} 
\eta^{D} &=& 1 - \frac{\ln\left[\frac{1}{4}\frac{1 + \theta(\alpha L_{A}/2)}{1 + \theta(2\alpha L_{A})} \left\{\frac{\theta(\alpha L_{A}/2)-\theta(\alpha L_{A})}{\theta(\alpha L_{A})
-\theta(2\alpha L_{A})}\right\}^{\theta(\alpha L_{A})}\right]}
{\ln\left[\frac{1}{4}\frac{1 + \theta(L_{A}/2)}{1 + \theta(2 L_{A})}\left\{\frac{\theta(L_{A}/2)-\theta(L_{A})}{\theta(L_{A})
-\theta(2 L_{A})}\right\}^{\theta(L_{A})}\right] }.\nonumber\\
\label{eq28}
\end{eqnarray}
Here, we have defined the function $\theta(L) = \sqrt{1 + (\lambda/2L)^{2}}$, with $\lambda = 2\pi \hbar/(m c)$ the
Compton wavelength. It is important to remark that in the non-relativistic limit $\lambda/L \rightarrow 0$, the expression
in Eq.(\ref{eq27}) reduces to the known Schr\"odinger limit
\begin{eqnarray}
\lim_{\lambda/L \rightarrow 0} \eta^{D} = 1 - 1/\alpha^{2}.
\label{eq29}
\end{eqnarray}
In the "ultra-relativistic" case of a massless Dirac particle, $\lambda/L \rightarrow \infty$, the efficiency converges to
the length independent limit
\begin{eqnarray}
\lim_{\lambda/L\rightarrow\infty} \eta^{D} = 1 - 1/\alpha.
\label{eq30}
\end{eqnarray}
The trend of the efficiency is shown in between both limits in Figure 3. It is worth to remark that,
since the expansion parameter $\alpha = L_{C}/L_{B}>1$, the efficiency in the strict non-relativistic
(Schr\"odinger) regime $\lambda/L\rightarrow 0$ Eq.(\ref{eq29}) is the highest possible one.
This is also clear from the asymptotics of the curves displayed in Fig. 3, where the "ultra-relativistic"
limit corresponding to massless Dirac particles ($\lambda/L\rightarrow\infty$) indeed represents
the less efficient regime for a fixed expansion parameter $\alpha$. 

\section{The Quantum Carnot cycle}

In this section, we shall discuss the quantum version of the Carnot cycle,
as applied to the statistical ensemble of Dirac single-particle systems under consideration. 
The thermodynamic cycle which
defines the corresponding heat engine is composed of four stages or trajectories
in Hilbert's space: Two isothermal and two iso-entropic
processes.
 
\begin{figure}[tbp]
\centering
\epsfig{file=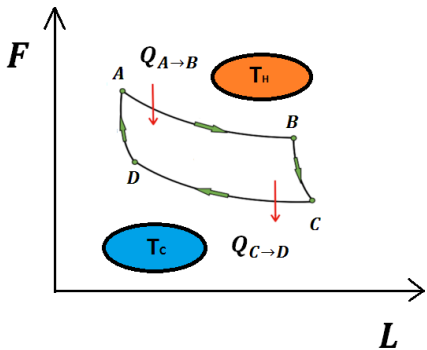,width=0.7\columnwidth,clip=}
\caption{(Color online)Pictorial description of the Force versus width of the potential well, for the Quantum Carnot Cycle described in the text.
\label{fig4}
}
\end{figure}

In the first stage, the system is brought into contact with a thermal reservoir
at temperature $T_{H}$. By keeping isothermal conditions, the width of the
potential well is expanded from $L_{A}\rightarrow L_{B}$. Since thermal equilibrium
with the reservoir is assumed along this process, the von Neumann entropy of the system 
achieves a maximum for the Boltzmann distribution \cite{vonNeumann,Tolman}
\begin{eqnarray}
p_{n}(L,\beta_{H}) = \left[Z(L,\beta_{H})\right]^{-1}e^{-\beta_{H}E_{n}^{D}(L)},
\label{eq31}
\end{eqnarray}
with $\beta = (k_{B}T)^{-1}$, and the normalization factor is given by the partition function
\begin{eqnarray}
Z(L,\beta) = \sum_{n=0}^{\infty} e^{-\beta E_{n}^{D}(L)} \sim \frac{2 L}{\lambda} e^{\beta mc^{2}} K_{1}(\beta m c^{2}).
\label{eq32}
\end{eqnarray}
Here, the second expression, as shown in Appendix, is the continuum approximation
to the discrete sum, valid in the physically relevant regime $\lambda \ll L$. Here, $K_{1}(x)$ is
a modified Bessel function of the second kind.

From a similar analysis as in the previous section, we conclude that the heat exchanged by the system
to
the thermal reservoir is given by
\begin{eqnarray}
Q_{A\rightarrow B} &=& \int_{L_{A}}^{L_{B}}\sum_{n=0}^{\infty}E_{n}(L)\frac{dp_{n}(L,\beta_{H})}{dL}dL
\nonumber\\
&=& -\frac{\partial\ln\left(\frac{Z(L_{B},\beta_{H})}{Z(L_{A},\beta_{H})} \right)}{\partial\beta_{H}}
+ \beta_{H}^{-1}\ln\left(\frac{Z(L_{B},\beta_{H})}{Z(L_{A},\beta_{H})} \right)\nonumber\\
&=& \beta_{H}^{-1}\ln\left(\frac{L_{B}}{L_{A}} \right).
\label{eq33}
\end{eqnarray}
In the second line, we have done integration by parts, and we made direct use of the definition Eq.(\ref{eq30})
of the partition function. The final result follows from substituting the explicit expression for
the partition function Eq.(\ref{eq32}).

Similarly, during the third stage of the cycle, the system is again brought into contact with a thermal
reservoir, but at a lower temperature $T_{C} < T_{H}$. Therefore, the probability distribution
of states in the ensemble is $p_{n}(L,\beta_{C})$, as defined in Eq.(\ref{eq31}), 
but with $T_{C}$ instead of $T_{H}$. The heat
released to the reservoir during this stage is given by the expression
\begin{eqnarray}
Q_{C\rightarrow D} 
= \beta_{C}^{-1}\ln\left(\frac{L_{D}}{L_{C}} \right).
\label{eq34}
\end{eqnarray}

The second and fourth stages of the cycle constitute iso-entropic trajectories. In order to analyze
these stages, we shall derive the "equation of state" for the statistical ensemble of single-particle systems.
When substituting the Boltzmann distribution $p_{n}(\beta,L)=\left[Z(\beta,L)\right]^{-1}\exp(-\beta E_{n}^{D}(L))$ into the expression for the von Neumann entropy Eq.(\ref{eq10}), we obtain the relation
\begin{eqnarray}
S = T^{-1} E + k_{B}\ln Z(\beta,L).
\label{eq35}
\end{eqnarray} 
Here, $E=\langle \hat{H} \rangle$ is the ensemble-average energy, as defined by Eq.(\ref{eq11}). The equation of state
is obtained from Eq.(\ref{eq35}) as
\begin{eqnarray}
F &=& -\left(\frac{\partial E}{\partial L}\right)_{S} = k_{B}T\frac{\partial}{\partial L}\ln Z(\beta,L)\nonumber\\
&=& \frac{k_{B}T}{L}.
\label{eq36}
\end{eqnarray}
In the last line, we have used the explicit analytical expression Eq.(\ref{eq32}) for the partition function to calculate
the derivative. The equation of state Eq.(\ref{eq36}) reflects that the ensemble of systems behaves as a one-dimensional ideal gas. This is not surprising, since the ensemble-average energy is given by
\begin{eqnarray}
E &=& \langle \hat{H} \rangle = -\frac{\partial}{\partial\beta}\ln Z(\beta,L)\nonumber\\
&=& m c^{2}\left(-1 - \frac{d}{dz}\ln K_{1}(z) \right),
\label{eq37}
\end{eqnarray}
where we have substituted explicitly the expression for the partition function, and in the final
step we defined $z = \beta\,m c^{2}$. Here, $K_{1}(x)$ is a modified Bessel function of the second kind. Eq.(\ref{eq37}) shows that the ensemble average energy
of the system is a function of the temperature solely, from which the ideal gas equation of state
follows as a natural consequence. We can thus define the "specific heat" at constant length, which
after Eq.(\ref{eq37}) is given by
\begin{eqnarray}
C_{L} &=& \left(\frac{\partial E}{\partial T} \right)_{L} = \frac{dE}{dT}= k_{B}z^{2}\frac{d^{2}}{dz^{2}}\ln K_{1}(z),
\label{eq38}
\end{eqnarray}
where $z=\beta m c^{2}$. It is interesting to remark that, based on the asymptotic behavior
of the modified Bessel functions of the second kind $K_{n}(z)\sim \sqrt{\pi/2}z^{-1/2}\exp(-z)+O(z^{-1})$,
the specific heat defined in Eq.(\ref{eq38}) presents the asymptotic limit $C_{L}\rightarrow k_{B}/2$
when $k_{B}T\ll m c^{2}$. This is the well known result for a classic non-relativistic
ideal gas in one dimension. This feature and the general temperature dependence of the ensemble
specific heat is displayed in Fig. 6.
The change in the ensemble averaged energy of the system, for a general process, is $dE = TdS - FdL$. Since
the ensemble-average energy is a function of temperature only, the differential equation
for an iso-entropic trajectory ($dS=0$) is\cite{note2}
\begin{eqnarray}
dE = C_{L}dT = -FdL = -k_{B}T\frac{dL}{L}.
\label{eq39}
\end{eqnarray}
Separating variables, after some algebra we obtain
\begin{eqnarray}
z \frac{d^{2}}{dz^{2}}\ln K_{1}(z) = \frac{dL}{L}.
\label{eq40}
\end{eqnarray}
Integrating Eq.(\ref{eq40}) between initial conditions $(z_{0},L_{0})$ and final
conditions $(z,L)$, we have
\begin{eqnarray}
\frac{L}{L_{0}} = \frac{K_{1}(z_{0})}{K_{1}(z)}e^{-z\frac{K_{0}(z)}{K_{1}(z)}
+ z_{0}\frac{K_{0}(z_{0})}{K_{1}(z_{0})}}.
\label{eq41}
\end{eqnarray}
Here, we made use of the identity for the modified Bessel function of the second kind  $K_{1}^{'}(z) = -K_{0}(z)-z^{-1}K_{1}(z)$,
with $z = \beta m c^{2}$. It is interesting to check that in the non-relativistic
limit $z\gg 1$, given the asymptotic
behavior of the modified Bessel function of the second kind $K_{n}(z)\sim \sqrt{\pi/2}z^{-1/2}\exp(-z)+O(z^{-1})$,
Eq.(\ref{eq41}) reduces to $L T^{1/2} = cnt.$ for the iso-entropic trajectory. On the other hand,
in the "ultra-relativistic" limit of a massless Dirac particle with $z\rightarrow 0$,
the iso-entropic trajectory is given by the equation $L T = cnt.$

We are now in conditions to discuss the second and fourth stages of the Carnot cycle.
The second stage of the process corresponds to an iso-entropic trajectory, parameterized in differential
and integral form by Eqs.(\ref{eq40}) and (\ref{eq41}), respectively. The work performed by the system during this process is given by
\begin{eqnarray}
W_{B\rightarrow C} &=&-\int_{L_{B}}^{L_{C}} F dL = -\int_{L_{B}}^{L_{C}} k_{B}T\frac{dL}{L}\nonumber\\
&=& -m c^{2}\int_{z_{H}}^{z_{C}}\frac{d^{2}}{dz^{2}}\ln K_{1}(z)dz
\label{eq42}
\end{eqnarray}
Here, in the second line we have used the differential equation defining the
iso-entropic trajectory, Eq.(\ref{eq40}).
Evaluating the integral in Eq.(\ref{eq42}), we explicitly obtain
\begin{eqnarray}
W_{B\rightarrow C} = k_{B}(T_{C} - T_{H})- m c^{2}\left[\frac{K_{0}(z_{C})}{K_{1}(z_{C})} -
\frac{K_{0}(z_{H})}{K_{1}(z_{H})}\right]
\label{eq43}
\end{eqnarray}
\begin{figure}[tbp]
\centering
\epsfig{file=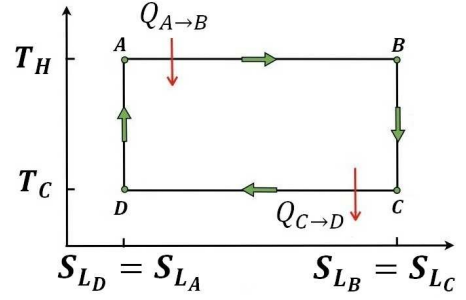,width=0.7\columnwidth,clip=}
\caption{(Color online)Pictorial description of the Force versus width of the potential well for the Quantum Carnot Cycle.
\label{fig5}
}
\end{figure}

The fourth and final stage of the cycle also corresponds to an iso-entropic trajectory $L_{D}\rightarrow L_{A}$, and the work performed
by the system against the external applied force is obtained similarly as in Eq.(\ref{eq42}),
\begin{eqnarray}
W_{D\rightarrow A} &=& -\int_{L_{D}}^{L_{A}}F dL\\
&=& k_{B}(T_{H} - T_{C})- m c^{2}\left[\frac{K_{0}(z_{H})}{K_{1}(z_{H})} -
\frac{K_{0}(z_{C})}{K_{1}(z_{C})}\right].\nonumber
\label{eq44}
\end{eqnarray}
Clearly, after Eqs.(\ref{eq43}) and (\ref{eq44}), we have $W_{B\rightarrow C} + W_{D\rightarrow A}=0$,
and hence the contribution of the work along the iso-entropic trajectories vanishes.

From the equation for the iso-entropic trajectory Eq.(\ref{eq41}), we conclude that the length ratios
are determined by the temperatures of the thermal reservoirs,
\begin{eqnarray}
\frac{L_{C}}{L_{B}} = \frac{L_{D}}{L_{A}} = \frac{K_{1}(z_{H})}{K_{1}(z_{C})}e^{-z_{C}\frac{K_{0}(z_{C})}{K_{1}(z_{C})}+ z_{H}\frac{K_{0}(z_{H})}{K_{1}(z_{H})}}.
\label{eq45}
\end{eqnarray}

\begin{figure}[tbp]
\centering
\epsfig{file=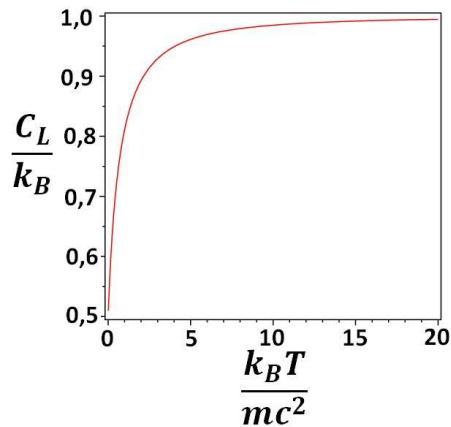,width=0.7\columnwidth,clip=}
\caption{(Color online)Specific heat of the statistical ensemble of single Dirac-particle systems, after Eq.(\ref{eq38}), as a function of temperature. 
\label{fig6}
}
\end{figure}

From Eq.(\ref{eq45}), we also obtain $L_{A}/L_{B} = L_{D}/L_{C}$. Substituting this relation
in the expression for the efficiency of the cycle, from Eq.(\ref{eq33}) and Eq.(\ref{eq34}) we have
\begin{eqnarray}
\eta^{C} &=& 1 - \frac{Q_{C\rightarrow D}}{Q_{A\rightarrow B}} = 1 - \frac{T_{C}}{T_{H}}\frac{\ln(L_{D}/L_{C})}
{\ln(L_{B}/L_{A})}\nonumber\\
&=& 1 - \frac{T_{C}}{T_{H}}.
\label{eq46}
\end{eqnarray}
Therefore, we have recovered the expression for the efficiency identical to the classical Carnot cycle.

\section{Conclusions}

By considering as a "working substance" the statistical ensemble for a Dirac single-particle system trapped in
a one-dimensional potential well, we have analyzed two different schemes for a
quantum heat engine. The first, that we have referred to as the Isoenergetic cycle, consists
of two iso-entropic and two isoenergetic trajectories. We obtained an explicit
expression for the efficiency of this cycle and showed that our analytical result, in the non-relativistic limit $\lambda/L \rightarrow 0$, reduces to the corresponding
one for a Schr\"odinger particle, as reported in the literature \cite{Bender_02}. Our results also indicate that
the efficiency for the Isoenergetic cycle is higher in the non-relativistic region of
parameters, that is for $L\gg \lambda$, when comparing at the same compressibility ratios $\alpha = L_{C}/L_{B}>1$. 
An exception is the case of massless Dirac particles, with $\lambda = \infty$, where it is not possible to achieve non-relativistic
conditions. This is of potential practical interest for graphene systems, where conduction electrons
are indeed described as massless chiral Dirac fermions\cite{Peres010,Castro_Neto09,Munoz010,Munoz012}. 

As a second candidate for a quantum heat engine, we discussed a version of the Carnot cycle,
composed by two iso-thermal and two iso-entropic trajectories. In order to achieve
iso-thermal conditions, we consider that the single-particle system is in thermal
equilibrium with macroscopic reservoirs at temperatures $T_{C} < T_{H}$, respectively. Therefore,
the statistical ensemble under consideration is described
by the density matrix $\hat{\rho} = e^{-\beta\hat{H}}/Z$, with $Z[\beta,L]={\rm{Tr}}e^{-\beta\hat{H}}$
the partition function.
We showed that the statistical properties of the ensemble are such that an equation of state
can be defined, as well as a specific heat, in analogy with a classical ideal gas in one dimension.
We obtained the equation for the iso-entropic trajectory, which in the non-relativistic
limit $k_{B}T \ll m c^{2}$ reduces to the classical result $L T^{1/2}=cnt$. On the other
hand, we also showed that in the "ultra-relativistic" limit of a massless Dirac particle,
as for instance conduction electrons in graphene, the iso-entropic trajectory is defined
by the equation $L T = cnt.$
We also showed that the efficiency for the Quantum Carnot cycle satisfies the same
relation that the classical one in terms of the temperatures of the thermostats,
that is $\eta = 1 - T_{C}/T_{H}$. Therefore, thermodynamics is remarkably robust: The Carnot limit
holds classically just as it does in the quantum regime, even in the quantum-relativistic limit.

Finally, we conclude by saying that the general statistical mechanical analysis presented in this work,
can in principle be applied to predict the thermodynamics of other one-dimensional single-particle systems of potential interest,
with different energy spectra. 

\section*{Acknowledgements}

The authors wish to thank Dr. P. Vargas, and Dr. E. Stockmeyer for interesting discussions.
E.M. acknowledges financial support from Fondecyt Grant 11100064. F.J.P.
acknowledges financial support from a Conicyt fellowship.

\section*{Appendix A}

We shall consider the physically meaningful eigenfunctions of the
self-adjoint extension \cite{Thaller,Carreau_90,Alonso_97,Alberto_96} of the {\it{free}} Dirac Hamiltonian 
\begin{eqnarray}
\hat{H}_{0} = -i\hbar c\,\bm{\alpha}\cdot\nabla + m c^{2}\hat{\beta},
\label{eq46b}
\end{eqnarray}
whose domain 
$\mathcal{D}(\hat{H}_{0})\subset \mathcal{H}_{\Omega}$
is a dense proper subset of the Hilbert space 
$\mathcal{H}_{\Omega}=L^{2}(\Omega)\oplus L^{2}(\Omega) \oplus L^{2}(\Omega)\oplus L^{2}(\Omega)\equiv L^{2}(\Omega,\mathbb{C}^4)$ of square-integrable (complex-valued) 4-component spinors in the closed
interval $x\in \Omega = [-L/2,L/2]$. It is convenient to
describe the four-component spinor $\hat{\psi}(x) = (\phi_{1},\phi_{2},\chi_{1},\chi_{2})$ as 
composed by a pair of two-component
spinors $\phi(x) = (\phi_{1},\phi_{2})$ and $\chi(x)=(\chi_{1},\chi_{2})$. In this notation,
$\phi(x)$ is denoted as the "large" component, while $\chi(x)$ is referred as the "small" component.

In general, the domain of $\hat{H}_{0}$ and its adjoint
$\hat{H}_{0}^{\dagger}$ verify  $\mathcal{D}(\hat{H}_{0})\subseteq \mathcal{D}(\hat{H}_{0}^{\dagger})$ \cite{Thaller}. On the other hand, physics requires for $\hat{H}_{0}$ to be self
adjoint. This condition is mathematically obtained 
by looking for a self-adjoint extension of $\hat{H}_{0}$, upon imposing appropriate boundary conditions
for the spinors at the boundary $\partial\Omega$ of the finite region $\Omega$. 
Therefore, for any pair of spinors in the domain
of the self-adjoint extension of the free Hamiltonian, we request 
for their inner product to satisfy
\begin{eqnarray}
\left(\hat{\psi}_{a},\hat{H}\hat{\psi}_{b}\right) &-& \left(\hat{H}\hat{\psi}_{a},\hat{\psi}_{b}\right)\nonumber\\
&\equiv& \int_{\Omega}\left[\hat{\psi}_{a}^{\dagger}\hat{H}\hat{\psi}_{b} - \left(\hat{H}\hat{\psi}_{a}\right)^{\dagger}\hat{\psi}_{b}\right]dV\nonumber\\
 &=& -i\hbar c\int_{\partial \Omega}
\hat{\mathbf{n}}\cdot\hat{\psi}_{a}^{\dagger}{\bm{\alpha}}\hat{\psi}_{b}dS = 0.
\label{eq47}
\end{eqnarray}
Here, $\hat{\mathbf{n}}$ is a unit vector normal to the surface $\partial\Omega$.
We have used integration by parts, and applied the generalized divergence theorem
to transform the volume integral over the region $\Omega$ into an integral over the 
boundary $\partial\Omega$. Notice
that, in particular, condition Eq.(\ref{eq47}) implies that for all
$\hat{\psi}(x)\in \mathcal{D}(\hat{H}_{0})$,  
\begin{eqnarray}
\left.\hat{\mathbf{n}}\cdot\hat{\psi}^{\dagger}c{\bm{\alpha}}\hat{\psi}\right|_{\partial\Omega}\equiv 
\left.\hat{\mathbf{n}}\cdot{\mathbf{j}}\right|_{\partial\Omega} = 0.
\label{eq48}
\end{eqnarray}
This
mathematical condition has a rather obvious physical interpretation, since 
the probability current vanishing at the boundary $\partial\Omega$ implies that
the particle is indeed "trapped" inside the finite region $\Omega$. We remark that this mathematical approach, with analogous boundary condition, has
been used in the past to model confinement of hadrons
in a finite region of space (so-called "bag model") \cite{Chodos2_74,Chodos_74}.

Therefore, the eigenvalue problem for the self-adjoint extension of the free Dirac Hamiltonian
in the one-dimensional region $\Omega = [-L/2,L/2]$,
\begin{eqnarray}
\hat{H}_{0}\hat{\psi} = E\hat{\psi}
\label{eq48b}
\end{eqnarray}
must be subject to the boundary conditions
\begin{eqnarray}
j^{1}(x=\pm L/2) = 0.
\label{eq49}
\end{eqnarray}

It is
convenient to write Eq.(\ref{eq48b}) as a pair of coupled differential equations
for each of the two components of the spinor,
\begin{eqnarray}
-i\hbar c\,\bm{\sigma}\cdot\nabla \chi + m c^{2}\phi &=& E \phi\nonumber\\
-i\hbar c\,\bm{\sigma}\cdot\nabla \phi - m c^{2}\chi &=& E \chi.
\label{eq50}
\end{eqnarray}

From Eq.(\ref{eq50}), we obtain
\begin{eqnarray}
\chi = \frac{-i\hbar c}{E + mc^{2}}\bm{\sigma}\cdot\nabla \phi.
\label{eq51}
\end{eqnarray}
This equation determines the "small" component $\chi$ of the spinor in terms of the "large" component
$\phi$. The last is obtained upon substitution of Eq.(\ref{eq51}) into Eq.(\ref{eq50}), as
the solution to the eigenvalue problem
\begin{eqnarray}
\bm{\sigma}\cdot\nabla\left(\bm{\sigma}\cdot\nabla\phi \right) = 
-\frac{E^{2}-m^{2}c^{4}}{\hbar^{2}c^{2}}\phi.
\label{eq52}
\end{eqnarray} 
For the one-dimensional region we are concerned $\Omega = [-L/2,L/2]$, we have $\phi = \phi(x)$,
and Eq.(\ref{eq52}) becomes
\begin{eqnarray}
\frac{d^{2}}{dx^{2}}\left(\begin{array}{c}\phi_{1}\\\phi_{2}\end{array}\right) + k^{2}\left(\begin{array}{c}\phi_{1}\\\phi_{2}\end{array}\right) = 0.
\label{eq53}
\end{eqnarray}
Here, we defined the wavenumber 
\begin{eqnarray}
k^2 \equiv (E^{2}-m^{2}c^{4})/(\hbar^{2}c^{2}).
\label{eq53b}
\end{eqnarray}
The general solution to Eq.(\ref{eq53}) is
\begin{eqnarray}
\phi(x) &=& A_{1}\left(\begin{array}{c}1\\0 \end{array}\right)e^{ikx} 
+ A_{2}\left(\begin{array}{c}0\\1\end{array} \right)e^{ikx}\nonumber\\
&+& B_{1}\left(\begin{array}{c}1\\0 \end{array}\right)e^{-ikx} 
+ B_{2}\left(\begin{array}{c}0\\1 \end{array}\right)e^{-ikx}.
\label{eq54}
\end{eqnarray}
We shall consider a spin-polarized particle, so we select the spin up solutions
by setting $A_{2}=0$, $B_{2}=0$. With this choice, from Eq.(\ref{eq51}) we obtain
$\chi$, which combined with Eq.(\ref{eq54}) gives the general solution
for the four-component spinor
\begin{eqnarray}
\hat{\psi}(x) 
= A_{1}\left(\begin{array}{c}1\\0\\0\\\frac{\hbar c k}{E + m c^2} \end{array}\right) e^{ikx}
+ B_{1}\left(\begin{array}{c}1\\0\\0\\\frac{-\hbar c k}{E + m c^2} \end{array}\right) e^{-ikx}.
\label{eq55}
\end{eqnarray}

By requesting for the condition of vanishing current at the boundaries Eq.(\ref{eq49}), we obtain
\begin{eqnarray}
j^{1}(x) &=& \hat{\psi}^{\dagger}(x)c\hat{\alpha}_{1}\hat{\psi}(x) \nonumber\\
&=& \frac{2\hbar c^{2} k}{E + m c^2}\left(|A_{1}|^2
- |B_{1}|^2 \right) = 0,
\label{eq56}
\end{eqnarray}
which implies $|A_{1}| = |B_{1}|$. We express this condition in the form
\begin{eqnarray}
B_{1} = A_{1}e^{i\theta},
\label{eq57}
\end{eqnarray}
with $\theta$ the relative phase between the two (complex) amplitudes. Therefore, we obtain
an entire family of (spin polarized) eigenfunctions for the self-adjoint extension of the
Dirac operator 
\begin{eqnarray}
\hat{\psi}(x) = A_{1}\left(\begin{array}{c}e^{i k x} + e^{-i k x + i\theta}\\0\\0\\
\frac{\hbar c k}{E + m c^2}\left[e^{i k x} - e^{-i k x + i\theta} \right]\end{array} \right)
\label{eq58}
\end{eqnarray}

Among all the possible phases $\theta$ in Eq.(\ref{eq58}), 
the appropriate physical
solution should converge to the solution of the equivalent Schr\"odinger problem 
in the non-relativistic limit $\hbar k \ll m c$. In this former case, the
Schr\"odinger (S) wave functions are of the form $\varphi_{n}^{S}(x) = A\sin(n\pi (x-L/2)/L)$,
with corresponding probability density $\rho^{S}(x)\propto \sin^{2}(n\pi (x - L/2)/L)$. Therefore,
we set the phase difference $\theta= k L + \pi$, 
\begin{eqnarray}
\hat{\psi}(x) = A\left(\begin{array}{c}\sin(k(x-L/2))\\0\\0\\
-\frac{i\hbar c k}{E + m c^2}\cos(k(x-L/2))\end{array} \right).
\label{eq59}
\end{eqnarray}
where we have defined $A\equiv 2 i A_{1}e^{ikL/2}$.
Clearly, the "small" component of the spinor in Eq.(\ref{eq59})
vanishes in the non-relativistic limit $\hbar k \ll m c$, thus
leading to a probability density $\rho(x) = \hat{\psi}^{\dagger}(x)\hat{\psi}(x) \propto \sin^{2}(k (x-L/2))$, in agreement with
the Schr\"odinger result. 

Eq.(\ref{eq59}) satisfies the condition $j^{1}(x) = j^{1}(\pm L/2) = 0$,
where the probability current is continuous and vanishes everywhere, in particular at the
boundaries of the confining region. 

A fundamental discrete
symmetry of the Dirac Hamiltonian is its invariance under parity \cite{Sakurai,Bjorken_Drell,Thaller} ($\hat{P}$: $x\rightarrow -x$, $p_{x}\rightarrow -p_{x}$), $[\hat{H}_{0},\hat{P}]=0$, which corresponds to a mirror spatial reflection, by leaving the spin direction invariant. 
It is straightforward to show that 
under parity, the spinor transforms as \cite{Bjorken_Drell,Sakurai,Thaller} 
$\hat{P}\hat{\psi}(x)\hat{P}^{-1} = e^{i\phi}\hat{\beta}\hat{\psi}(-x)$. 
On the other hand, the probability density defined as $\rho(x) = \hat{\psi}^{\dagger}(x)\hat{\psi}(x) = c^{-1}j^{0}(x)$ is the time-component of the
Lorentz 4-vector $j^{\mu} = c\bar{\psi}(x)\gamma^{\mu}\hat{\psi}(x)$, with $\bar{\psi}(x) \equiv \gamma^{0}\hat{\psi}^{\dagger}(x)$, and where the explicit covariant notation $\gamma^{i}=\hat{\beta}\hat{\alpha}_{i}$ (i=1,2,3), $\gamma^{0} = \hat{\beta}$ was invoked. Under parity,
the space components of a Lorentz 4-vector change sign, whereas the time component remains invariant \cite{Bjorken_Drell,Sakurai,Thaller}, and hence $\rho(-x) = \rho(x)$. Therefore,
we demand
for the eigenfunctions of the self-adjoint extension of the Dirac Hamiltonian to possess
this inversion symmetry,
\begin{eqnarray}
\rho(x) = \rho(-x),&\forall x \in [-L/2,L/2].
\label{eq60}
\end{eqnarray}
From Eq.(\ref{eq59}), and the definition of the probability density
\begin{eqnarray}
\rho(\pm x) &=& \hat{\psi}^{\dagger}(\pm x)\hat{\psi}(\pm x)\nonumber\\
&=& 4 |A_{1}|^2\left[1 - \left(1 - \frac{\hbar^2 c^2 k^2}{(E+m c^2)^2} \right)\right.\nonumber\\
&&\left.\times\cos^2(k(\pm x - L/2)) \right]
\label{eq61}
\end{eqnarray}
By using the trigonometric identity $\cos^2(\alpha) = (1 + \cos(2\alpha))/2$, we obtain that
the condition for Eq.(\ref{eq60}) to be satisfied is
\begin{eqnarray}
\cos(kL + 2 k x) = \cos(kL - 2 k x),&\forall x \in [-L/2,L/2]
\label{eq62}
\end{eqnarray}
This is further simplified to yield the equation
\begin{eqnarray}
2\sin(2 k x)\sin(k L) = 0,&\forall x \in [-L/2,L/2],
\label{eq63}
\end{eqnarray}
whose solutions are
\begin{eqnarray}
k \equiv k_{n} = \frac{n\pi}{L},& n\in\mathbb{N}.
\label{eq64}
\end{eqnarray}
Therefore, the spectrum of the self-adjoint extension of
the Dirac Hamiltonian representing a single particle trapped in the one-dimensional
infinite potential well is discrete, and
given by
\begin{eqnarray}
E_{n} = \sqrt{c^{2}\hbar^{2}k_{n}^{2} + m^{2}c^{4}} 
= \sqrt{\left(\frac{\hbar c n\pi}{L}\right)^{2} + m^{2}c^{4}}.
\label{eq65}
\end{eqnarray}
By introducing the definition of the Compton wavelength $\lambda = 2\pi\hbar/(mc)$, the spinor-eigenfunction Eq.(\ref{eq59}),
with $k_{n}$ quantized by Eq.(\ref{eq64}), can be written
as Eq.(\ref{eq4}) in the main text. By subtracting the rest
energy from Eq.(\ref{eq65}), one obtains Eq.(\ref{eq5})
for the Dirac particle spectrum. 

\section*{Appendix B}

We here discuss the continuum approximation to the discrete partition function Eq.(\ref{eq30}).
\begin{eqnarray}
Z(L,\beta) = \sum_{n=0}^{\infty}e^{-\beta m c^{2}\left(\sqrt{1 + \left(\frac{n\lambda}{2L} \right)^{2}}-1\right)}\nonumber
\end{eqnarray}
Here, let us define the discrete variable $x_{n} = n\lambda/(2L)$. The spacing between two consecutive
values is given by $\Delta x = x_{n+1} - x_{n} = \lambda/(2 L)$, and hence the
expression for the partition function can be written as
\begin{eqnarray}
Z(L,\beta) = \frac{2 L}{\lambda}\sum_{n=0}^{\infty}\Delta x\,e^{-\beta m c^{2}\left(\sqrt{1+x_{n}^{2}}-1 \right)}
\nonumber
\end{eqnarray}
For physically relevant sizes of the potential well, we shall have $\lambda/L \ll 1$, which allow us to
take the continuum limit in the sense of a Riemann sum for the previous equation, 
\begin{eqnarray}
Z(L,\beta) &&\rightarrow \frac{2 L}{\lambda} \int_{0}^{\infty} dx e^{-\beta m c^{2}\left(\sqrt{1 + x^{2}}-1 \right)}\nonumber\\
&=&\frac{2 L}{\lambda} \int_{0}^{\infty}dt e^{-\beta m c^{2}\left(\cosh(t)-1 \right)}\cosh(t)\nonumber\\
&=&\frac{2 L}{\lambda} e^{\beta m c^{2}} K_{1}\left(\beta m c^{2} \right).\nonumber
\end{eqnarray}
Here, in the second line we have made the substitution $x=\sinh(t)$, and $K_{1}(x)$ is a modified Bessel function
of the second kind.
\bibliographystyle{apsrev}

\begin{thebibliography}{27}
\bibitem[{not({\natexlab{a}})}]{note1}
\bibinfo{note}{A necessary condition for entropy to remain constant is
  $\sum_{n=1}^{\infty}(1 + \ln(p_{n}(L))dp_{n}/dL=0$. This is clearly less
  stringent than the sufficient condition $dp_{n}/dL = 0$ for all n.}
\bibitem[{not({\natexlab{b}})}]{note2}
\bibinfo{note}{The necessary condition for an iso-entropic trajectory in this
  case differs from the more stringent sufficient condition $dp_{n}/dL = 0$
  requested in the E-Carnot cycle (see note 18). In terms of the distribution
  $\{p_{n}(L,\beta)\}$, a trajectory with $dS=0$ is defined by
  $\sum_{n=1}^{\infty}[1 + \ln(p_{n}(L))]dp_{n}/dL=0$. This implies that
  $\beta$ must change as a function of $L$ along the process in order to
  satisfy the iso-entropic constraint.}
\end{thebibliography}

\end{document}